\documentclass[11pt,floatfix,showpacs]{revtex4}
\usepackage[dvips]{graphicx,color}
\usepackage{amsmath}
\usepackage{amsfonts}
\usepackage[latin1]{inputenc}

\begin{document}

\title{Nonlocal Polyakov-Nambu-Jona-Lasinio model with wavefunction renormalization at finite temperature and chemical potential}

\author{G.A.Contrera$^{a,b}$, M.Orsaria$^{b,c}$ and
N.N.Scoccola$^{a,b,d}$}

\address{
$^{a}$ Physics Department, Comisi\'on Nacional de
Energ\'{\i}a Atómica, Av.Libertador 8250, 1429 Buenos Aires,
Argentina\\
$^{b}$ CONICET, Rivadavia 1917, 1033 Buenos Aires, Argentina\\
$^{c}$
Gravitation, Astrophysics and Cosmology Group, FCAyG, UNLP, La
 Plata, Argentina,\\
$^{d}$ Universidad Favaloro, Sol{\'\i}s 453, 1078 Buenos Aires,
Argentina}

\begin{abstract}
We study the phase diagram of strongly interacting matter in the
framework of a non-local SU(2) chiral quark model which includes
wave function renormalization and coupling to the Polyakov loop.
Both non-local interactions based on the frequently used
exponential form factor, and on fits to the quark mass and
renormalization functions obtained in lattice calculations are
considered. Special attention is paid to the determination of the
critical points, both in the chiral limit and at finite quark
mass. In particular, we study the position of the Critical End
Point as well as the value of the associated critical exponents
for different model parameterizations.
\end{abstract}


\maketitle

\section{Introduction}
At low temperatures and densities strongly interacting matter is
believed to be in a phase in which chiral symmetry is broken and
the quarks are confined. However, as the temperature ($T$) and/or
the chemical potential ($\mu$) increase some kind of transition to
a chiral restored and/or deconfined phase is expected to happen.
The detailed understanding of this phenomenon has become an issue
of great interest in recent years, both theoretically and
experimentally\cite{Rischke:2003mt}. From the theoretical point of view, even if a
significant progress has been made in the development of ab initio
calculations such as lattice QCD \cite{All03,Fod04,Kar03}, these
are not yet able to provide a full understanding of the QCD phase
diagram, due to the well-known difficulties of dealing with small
current quark masses and finite chemical potentials. Thus, it is
important to develop effective models that show consistency with
lattice results and can be extrapolated into regions not
accessible by lattice calculation techniques. Among them, the
local Nambu-Jona-Lasinio (NJL) has been widely used to describe
the behavior of strongly interacting matter at finite temperature
and density \cite{reports}. In recent years an extension of the
NJL model has been proposed in which the coupling of the quarks to
the Polyakov loop is included. This so-called
Polyakov-Nambu-Jona-Lasinio (PNJL) model
\cite{Meisinger:1995ih,Fukushima:2003fw,Megias:2004hj,Ratti:2005jh,Roessner:2006,Mukherjee:2006hq,Sasaki:2006ww}
allows to study the chiral and deconfinement transitions in a
common framework. As an improvement over local models, the study
of the phase diagram of chiral quark models that include nonlocal
interactions \cite{Rip97} has been undertaken
\cite{GDS00,Tum:2005,DGS04}. These theories can be viewed as
nonlocal extensions of the NJL model. In fact, nonlocality arises
naturally in the context of several successful approaches to
low-energy quark dynamics as, for example, the instanton liquid
model \cite{Schafer:1996wv} and the Schwinger-Dyson resummation
techniques \cite{RW94}. Lattice QCD calculations
\cite{Parappilly:2005ei,Furui:2006ks} also indicate that quark
interactions should act over a certain range in momentum space. In
addition, several studies
\cite{BB95,BGR02,Scarpettini:2003fj,GomezDumm:2006vz} have shown
that nonlocal chiral quark models provide a satisfactory
description of hadron properties at zero temperature and density.
The aim of the present work is to extend previous studies
of the chiral and
deconfinement transitions in the framework of non-local chiral
models with coupling to the Polyakov
loop\cite{Contrera:2007wu,Blaschke:2007np,Hell:2008cc} by considering more
general quark interactions. Following Refs.
\cite{Noguera:2005ej,Noguera:2008}, we will adopt as the basic
ingredient a reliable description of the $T=\mu=0$ quark
propagator as given from fundamental studies, such as lattice QCD.
In this sense, it should be noticed that most of the finite $T$
and/or $\mu$ calculations performed so far in the context of
non-local chiral quark models have used exponential regulators and
neglected the wave function renormalization (WFR) in the quark
propagator. Recent lattice QCD calculations suggest, however, that
the WFR can be of the order of 30 \% (or even more) at zero
momentum \cite{Parappilly:2005ei,Furui:2006ks}. Moreover, these
calculations also show that the quark masses tend to their
asymptotic values in a rather soft way. Thus, it is important to
perform a detailed study of the impact of the incorporation of
these features on the predictions for the phase diagram and
associated quantities. The lagrangian we will use is the minimal
extension which allows to incorporate the full momentum dependence
of the quark propagator, through its mass and wave function
renormalization. Using such a model we will investigate the phase
diagrams corresponding to different parameterizations, including
one based on fits to the quark mass and renormalization functions
obtained in lattice calculations, both in the chiral limit and for
finite quark mass. The position of the critical points as well as
the value of the associated critical exponents will be also
studied.

This article is organized as follows. In Sec. II we provide a
description of the model and its parameterizations. In Sec. III we
present and discuss the results obtained in the chiral limit,
while those corresponding to finite values of the quark mass are
given and analyzed in Sec. IV. In Sec. V we present a summary of
our main results and conclusions. Finally, we include two
appendices. In App. A we provide some details concerning the
derivation of the Landau expansion associated with our model in
the chiral limit, while in App. B we describe the formalism used
to determine the position of the Critical End Point.

\section{The model and its parameterizations}

We consider a nonlocal SU(2) chiral quark model which
includes quark couplings to the color gauge fields. The corresponding
Euclidean effective action is given by
\begin{equation}
S_{E}= \int d^{4}x\ \left\{
\bar{\psi}(x)\left( -i\gamma_{\mu}D_{\mu}
+\hat{m}\right)  \psi(x) -\frac{G_{S}}{2} \Big[ j_{a}(x)j_{a}(x)- j_{P}%
(x)j_{P}(x)\Big]+ \ {\cal U}\,(\Phi[A(x)])\right\}  \ , \label{action}%
\end{equation}
where $\psi$ is the $N_{f}=2$ fermion doublet $\psi\equiv(u,d)^T$,
and $\hat{m}=diag(m_{u},m_{d})$ is the current quark mass matrix.
In what follows we consider isospin symmetry, that is
$m=m_{u}=m_{d}$. The fermion kinetic term includes a covariant
derivative $D_\mu\equiv \partial_\mu - iA_\mu$, where $A_\mu$ are
color gauge fields, and the operator $\gamma_\mu\partial_\mu$ in
Euclidean space is defined as $\vec \gamma \cdot
\vec \nabla + \gamma_4\frac{\partial}{\partial
\tau}$, with $\gamma_4=i\gamma_0$. The nonlocal currents
$j_{a}(x),j_{P}(x)$ are given by
\begin{align}
j_{a}(x)  &  =\int d^{4}z\ g(z)\ \bar{\psi}\left(  x+\frac{z}{2}\right)
\ \Gamma_{a}\ \psi\left(  x-\frac{z}{2}\right)  \ ,\nonumber\\
j_{P}(x)  &  =\int d^{4}z\ f(z)\ \bar{\psi}\left(  x+\frac{z}{2}\right)
\ \frac{i {\overleftrightarrow{\rlap/\partial}}}{2\ \kappa_{p}}
\ \psi\left(  x-\frac{z}{2}\right) . \label{currents}%
\end{align}

Here, $\Gamma_{a}=(\leavevmode\hbox{\small1\kern-3.8pt\normalsize1},i\gamma
_{5}\vec{\tau})$ and $u(x^{\prime}){\overleftrightarrow{\partial}%
}v(x)=u(x^{\prime})\partial_{x}v(x)-\partial_{x^{\prime}}u(x^{\prime})v(x)$.
The functions $g(z)$ and $f(z)$ in Eq.(\ref{currents}), are
nonlocal covariant form factors characterizing the corresponding
interactions. The four standard quark currents $j_{a}(x)$ require
the same $g(z)$ form factor to guarantee chiral invariance. The
term $\frac{G}{2}j_{P}(x)j_{P}(x)$ is self-invariant under chiral
transformations. The scalar-isoscalar component of the $j_{a}(x)$
current will generate the momentum dependent quark mass in the
quark propagator, while the ``momentum'' current, $j_{P}(x),$ will
be responsible for a momentum dependent wave function
renormalization of this propagator. For convenience, we take the
same coupling parameter, $G_{S},$ for both interaction terms.
Note, however, that the relative strength between both interaction
terms will be controlled by the mass parameter $\kappa_{p}$
introduced in Eq.(\ref{currents}). In what follows it is
convenient to Fourier transform $g(z)$ and $f(z)$ into momentum
space. Note that Lorentz invariance implies that the Fourier
transforms $g(p)$ and $f(p)$ can only be functions of~$p^{2}$.

To proceed we perform a standard bosonization of the
theory. Thus, we introduce the bosonic fields $\sigma_{1,2}(x)$
and $\pi_a(x)$, and integrate out the quark fields. In what
follows, we work within the mean-field approximation (MFA), in
which these bosonic fields are replaced by their vacuum
expectation values $\sigma_{1,2}$ and  $\pi_a = 0$.
Since we are interested in studying the characteristics of the
chiral phase transition we have to extend the so obtained
bosonized effective action to finite temperature $T$ and chemical
potential $\mu$. In the present work this is done by using the
Matsubara formalism. Concerning the gluon fields we will assume
that they provide a constant background color field $A_4 = i A_0 =
i g\,\delta_{\mu 0}\, G^\mu_a \lambda^a/2$, where $G^\mu_a$ are
the SU(3) color gauge fields. Then the traced Polyakov loop, which
is taken as order parameter of confinement, is given by
$\Phi=\frac{1}{3} {\rm Tr}\, \exp( i \phi/T)$, where $\phi = i
A_0$. We will work in the so-called Polyakov gauge, in which the
matrix $\phi$ is given a diagonal representation $\phi = \phi_3
\lambda_3 + \phi_8 \lambda_8$. This leaves only two independent
variables, $\phi_3$ and $\phi_8$. At vanishing chemical potential,
owing to the charge conjugation
properties of the QCD Lagrangian, the mean field traced Polyakov loop
is expected to be a real quantity. Since $\phi_3$ and $\phi_8$ have
to be real-valued~\cite{Rossner:2007ik}, this condition implies $\phi_8 =
0$. In general, this need not be the case at finite
$\mu$ \cite{Elze:1987,Dumitru:2005ng,Fukushima:2006uv}. As in e.g.
Refs.\cite{Roessner:2006,Abuki:2008ht,Rossner:2007ik,GomezDumm:2008sk} we will assume
that the potential $\cal U$ is such that the condition $\phi_8=0$ is well
satisfied for the range of values of $\mu$ and $T$ investigated here. The
mean field traced Polyakov loop is then given by
$\Phi = \Phi^\ast = [ 1 + 2\,\cos (\phi_3/T) ]/3$.

Within this framework the mean field thermodynamical potential $\Omega^{\rm MFA}$ results
\begin{align}
\Omega^{\rm MFA} =  \,- \,\frac{4 T}{\pi^2} \sum_{c} \int_{p,n}
\mbox{ln} \left[ \frac{ (\rho_{n, \vec{p}}^c)^2 +
M^2(\rho_{n,\vec{p}}^c)}{Z^2(\rho_{n, \vec{p}}^c)}\right]+
\frac{\sigma_1^2 + \kappa_p^2\ \sigma_2^2}{2\,G_S} +
{\cal{U}}(\Phi ,T). \label{granp}
\end{align}
Here, the shorthand notation $\int_{p,n} = \sum_n \int d^3\vec p/(2\pi)^3$ has been used, and $M(p)$ and $Z(p)$
are given by
\begin{eqnarray}
M(p) & = & Z(p) \left[m + \sigma_1 \ g(p) \right] ,\nonumber\\
Z(p) & = & \left[ 1 - \sigma_2 \ f(p) \right]^{-1}.
\label{mz}
\end{eqnarray}
In addition,
we have defined
\begin{equation}
\Big({\rho_{n,\vec{p}}^c} \Big)^2 =
\Big[ (2 n +1 )\pi  T- i \mu + \phi_c \Big]^2 + {\vec{p}}\ \! ^2 \ ,
\end{equation}
where the quantities $\phi_c$  are given by the relation $\phi =
{\rm diag}(\phi_r,\phi_g,\phi_b)$. Namely, $\phi_c = c \ \phi_3$
with $c = 1,-1,0$ for $r,g,b$, respectively.

To proceed we need to specify the explicit form of the Polyakov loop effective potential.
Following Ref. \cite{Roessner:2006} we consider
\begin{equation}
{\cal{U}}(\Phi ,T) = \left[-\,\frac{1}{2}\, a(T)\,\Phi^2 \;+\;b(T)\, \ln(1
- 6\, \Phi^2 + 8\, \Phi^3 - 3\, \Phi^4)\right] T^4 \ ,
\end{equation}
where the coefficients are parametrized as
\begin{equation}
a(T) = a_0 +a_1 \left(\dfrac{T_0}{T}\right) + a_2\left(\dfrac{T_0}{T}\right)^2
\qquad ;
\qquad
b(T) = b_3\left(\dfrac{T_0}{T}\right)^3,
\end{equation}
and the values of $T_0$, $a_i$ and $b_3$ are fitted to QCD lattice
results.

$\Omega^{\rm MFA}$ turns out to be divergent and, thus, needs to be regularized.
For this purpose we use the same prescription as in e.g. Ref. \cite{Tum:2005}. Namely,
\begin{equation}
\Omega^{\rm MFA}_{reg} = \Omega^{\rm MFA} - \Omega^{free} + \Omega^{free}_{reg} + \Omega_0,
\label{omegareg}
\end{equation}
where $\Omega^{free}$ is obtained from Eq.(\ref{granp}) by setting
$\sigma_1 = \sigma_2=0$ and
$\Omega^{free}_{reg}$ is the regularized expression for the
quark thermodynamical potential in the absence of fermion interactions,
\begin{equation}
\Omega^{free}_{reg} = -4\ T \int \frac{d^3 \vec{p}}{(2\pi)^3}\;
\sum_{c}
\left[ \ln\left( 1 + \exp\left[-\frac{E_p -\mu + i \phi_c}{T}\right] \right) +
       \ln\left( 1 + \exp\left[-\frac{E_p +\mu + i \phi_c}{T}\right] \right)
       \right],
\label{freeomegareg}
\end{equation}
with $E_p = \sqrt{\vec{p}^2+m^2}$. Finally, note that in
Eq.(\ref{omegareg}) we have included a constant $\Omega_0$ which
is fixed by the condition that $\Omega^{\rm MFA}_{reg}$ vanishes
at $T=\mu=0$.

The mean field values $\sigma_{1,2}$ and $\Phi$ at a given temperature or
chemical potential, are obtained from a set of three coupled ``gap'' equations.
This set of equations follows from the minimization of the regularized thermodynamical
potential, that is
\begin{equation}
\frac{\partial\Omega^{\rm MFA}_{reg}}{\partial\sigma_{1}} =
\frac{\partial\Omega^{\rm MFA}_{reg}}{\partial\sigma_{2}}=
\frac{\partial\Omega^{\rm MFA}_{reg}}{\partial\Phi}=0.
\label{fullgeq}
\end{equation}

Once the mean field values are obtained, the behavior
of other relevant quantities as a function of
temperature and chemical potential can be
determined. Here, we will be particularly interested in the chiral
quark condensate $\langle\bar{q}q\rangle$ and the quark density
$\rho$ defined by
\begin{eqnarray}
\langle \bar q q\rangle  =
\frac{\partial\Omega^{\rm MFA}_{reg}}{\partial m}
\qquad ; \qquad
\rho  =
- \frac{\partial\Omega^{\rm MFA}_{reg}}{\partial \mu} \ ,
\label{cond}
\end{eqnarray}
as well as their corresponding susceptibilities, i.e. the chiral susceptibility $\chi_{ch}$ and
the quark number susceptibility $\chi_q$, defined by
\begin{eqnarray}
\chi_{ch}  =  \frac{\partial\,\langle\bar qq\rangle}{\partial m}
\qquad ; \qquad
\chi_q  =  \frac{\partial \rho}{\partial \mu} \ .
\end{eqnarray}
Finally, the specific heat $C_{V}$, is expressed as
\begin{eqnarray}
C_{V} =  - T\ \frac{\partial^2\Omega^{\rm MFA}_{reg}}{\partial T^2} \ .
\label{spheat}
\end{eqnarray}

In order to fully specify the model under consideration we have to fix the model
parameters as well as the form factors $g(q)$ and $f(q)$ which characterize the
non-local interactions. Here, following Ref. \cite{Noguera:2008} we consider two different types
of functional dependencies for these form factors. The first one corresponds to
the often used exponential forms,
\begin{equation}
g(q)= \mbox{exp}\left(-q^{2}/\Lambda_{0}^{2}\right) \qquad ;
\qquad f(q)= \mbox{exp}\left(-q^{2}/\Lambda_{1}^{2}\right).
\label{regulators}
\end{equation}
Note that the range (in momentum space) of the nonlocality in each channel is determined by
the parameters $\Lambda_0$ and $\Lambda_1$, respectively. Fixing the current quark mass
and chiral quark condensate at $T=\mu=0$ to the reasonable values $m = 5.7$ MeV and
$\langle\bar{q}q\rangle^{1/3} = 240$ MeV, the rest of the parameters are determined so as to reproduce
the empirical values $f_\pi = 92.4$ MeV and
$m_\pi = 139$ MeV, and $Z(0) = 0.7$ which is within the range of values suggested by recent lattice
calculations[8, 10]. In what follows this choice of model parameters and
form factors will be referred as parametrization Set B.
The second type of form factor functional forms we consider is given by
\begin{eqnarray}
g(q)  = \frac{1+\alpha_z}{1+\alpha_z\ f_z(q)} \frac{\alpha_m \ f_m (q) -m\ \alpha_z f_z(q)}
{\alpha_m - m \ \alpha_z }
\qquad ; \qquad
f(q)  = \frac{ 1+ \alpha_z}{1+\alpha_z \ f_z(q)} f_z(q)\ ,
\label{regulators_set2}
\end{eqnarray}
where
\begin{equation}
f_{m}(q) = \left[ 1+ \left( q^{2}/\Lambda_{0}^{2}\right)^{3/2} \right]^{-1}
\qquad , \qquad
f_{z}(q) = \left[ 1+ \left( q^{2}/\Lambda_{1}^{2}\right) \right]^{-5/2}.
\label{parametrization_set2}
\end{equation}
As shown in Ref. \cite{Noguera:2008}, taking $m = 2.37$ MeV,
$\alpha_m = 309$ MeV, $\alpha_{z}=-0.3$, $\Lambda_0 = 850$ MeV and
$\Lambda_1 = 1400$~MeV one can very well reproduce the momentum
dependence of mass and renormalization functions obtained in
lattice calculations as well as the physical values of $m_\pi$ and
$f_\pi$. In what follows this choice of model parameters and form
factors will be referred as parametrization Set~C. Finally, in
order to compare with previous studies where the wave function
renormalization of the quark propagator has been ignored we
consider a third parametrization (Set A). In such a case we take
$Z(p)$ = 1 (i.e. $f(p)=0$) and an exponential parametrization for
$g(p)$. Such a model corresponds to the ``Scheme II'' discussed in
Ref.\cite{GomezDumm:2006vz}, from where we take the parameters
corresponding to $\langle\bar{q}q\rangle^{1/3} = 240$ MeV. The
values of the model parameters for each of the chosen
parameterizations are summarized in Table I.

\section{Phase diagram in the chiral limit}

In order to investigate the details of the phase diagram of the non-local models under study it is convenient
to consider first the chiral limit $m=0$. In this limit, general considerations imply
that for sufficiently small values of chemical potential the chiral restoration transition is
of second order with the transition temperature $T_c$
decreasing as $\mu$ increases. At a certain value of $\mu=\mu_{TCP}$ the transition becomes of first order.
The point in the $T-\mu$ plane defined by $(T_{TCP},\mu_{TCP})$ corresponds to the so-called ``tricritical point" (TCP).
For values of $\mu > \mu_{TCP}$
the corresponding $T_c$ continues to decrease until it reaches zero. This marks the end of the critical line,
$\mu_c(0)$ being the corresponding critical chemical potential.

In the following we will concentrate on the second order transition region. In such a region, for a given chemical
potential $\mu$, the condensate $\langle\bar{q}q\rangle$ goes to zero when the temperature $T$ approaches from below
the critical value $T_c(µ)$, above which $\langle\bar{q}q\rangle=0$ and the chiral symmetry is restored. Thus,
for $T \sim T_c(µ)$ the thermodynamical potential admits an expansion in powers of the order parameter (in
this case the quark condensate). As discussed in detail in App. A, in the chiral limit such expansion reads
\begin{eqnarray}
\Omega_{reg}^{\rm MFA}
& = &
\hat\Omega(\mu, T,\Phi_c, \sigma_{2c})  +
A_c \ \langle \bar q q\rangle^2  +
C_c \ \langle \bar q q\rangle^4 + {\cal O}\left(\langle \bar q q\rangle^6\right),
\label{expansion2}
\end{eqnarray}
where the explicit expressions of $\hat\Omega$ and the coefficients $A_c$ and $C_c$ are given in Eqs.(\ref{aa})
and (\ref{ab}), respectively.
Having established the Landau expansion in terms of the chiral condensate as single independent variable, we can now
analyze the characteristics of the phase transition following the standard textbook methods. For $C_c > 0$ the
system undergoes a second order phase transition at a critical temperature $T_c$. For each value of $\mu$, this critical
temperature can be obtained by solving a set of coupled equations given by the condition $A_c=0$ supplemented by
Eqs.(\ref{othereqs}). The values of $T_c(\mu)$ so obtained define a second order transition curve in the $(T,\mu)$ plane.
As already mentioned, such a curve is a decreasing function of $\mu$ which starts at the critical temperature corresponding to
vanishing chemical potential $T_c(0)$ and ends up at the tricritical point. The position of TCP can be determined
by imposing the additional condition $C_c=0$. Namely, to obtain it one has to solve the set of coupled equations
given by $A_c= C_c=0$ together with Eqs.(\ref{othereqs}).

To analyze the critical line beyond the TCP it is convenient to take $T$ as
independent variable and consider $\mu_c(T)$. For $T < T_{TCP}$, the transition turns out to be discontinuous (i.e. first order).
In this case, for each value of $T$, there is a region of values of $\mu$ for which three different solutions of the full
gap equations,
Eqs.(\ref{fullgeq}), exist. Two of them correspond to minima of the grand potential and the third one to a maximum.
In the chiral limit considered in the present section, one of the minima has $\sigma_1=0$, while
in the other $\sigma_1$ takes a
finite (in general, non-negligible) value.
Then, $\mu_c$ corresponds to the chemical potential at which the pressure associated with these two minima coincide.

\begin{figure}[hbt]
\includegraphics[width=0.9 \textwidth]{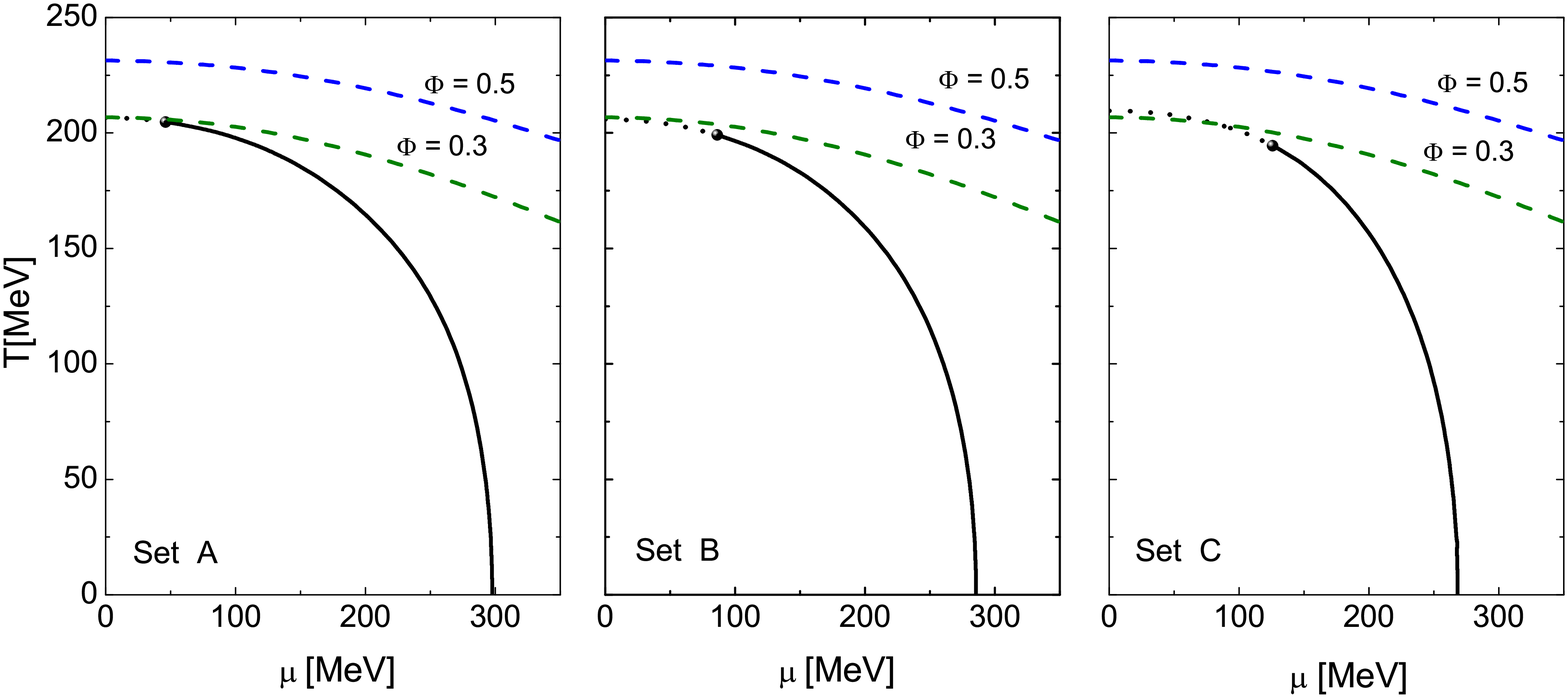}
\caption{Phase diagrams in the chiral limit for the three parameterizations
considered. Set B and Set C include quark wave function renormalization
while Set A does not. Set A and Set B correspond to exponential form
factors while Set C to lattice motivated form factors. The dotted
line corresponds to the second order chiral transition and
the full line to that of first order one. The dashed
lines correspond to the deconfinement transition (the lower and
upper lines being for $\Phi = 0.3$ and $\Phi = 0.5$,
respectively). } \label{figPhDiagCh}
\end{figure}

The phase diagrams corresponding to our three parameterizations
are displayed in Fig.\ref{figPhDiagCh} while the position of the
characteristic points are given in Table \ref{tab2}. In
Fig.\ref{figPhDiagCh} the dotted line indicates the second order
chiral transition line, the full line that of first order and the
dashed lines correspond to the deconfinement transition (the lower
and upper lines correspond to $\Phi = 0.3$ and $\Phi =
0.5$, respectively). We see that as $\mu$ increases there appears
a region where the system remains in its confined phase (signalled by
$\Phi$ smaller than $\simeq 0.3$) even though chiral symmetry
has been restored. This corresponds to the recently proposed
quarkyonic phase \cite{McLerran:2007qj}. We observe that the
general shape of the three diagrams is very similar with values of
the critical temperatures at $\mu=0$ differing by less than 4 MeV.
In the case of the critical chemical potential at $T=0$ the
difference between the three sets is somewhat larger. Comparing
the result of Set A with that of Set B we see that the inclusion of the
wave function renormalization implies a decrease of about 10 MeV
in the value of $\mu_c(0)$. The use of the softer form factors
involved in the lattice inspired parametrization Set C leads to a
further decrease of $\sim 10$ MeV . The feature of the phase
diagram that turns out to be most sensitive to the model
parametrization is the position of the TCP. In fact, although
the three values of {\bf $T_{TCP}$} are in a range of about 10
MeV, the value of $\mu_{TCP}$ increases by about a factor 2 when
the wave function renormalization is included (i.e. when one goes
from Set A to Set B) and by an extra factor $\sim 3/2$ when the lattice
inspired parametrization Set C is used (i.e. when one goes from Set B to
Set C).

\begin{figure}[hbt]
\includegraphics[width=0.9\textwidth]{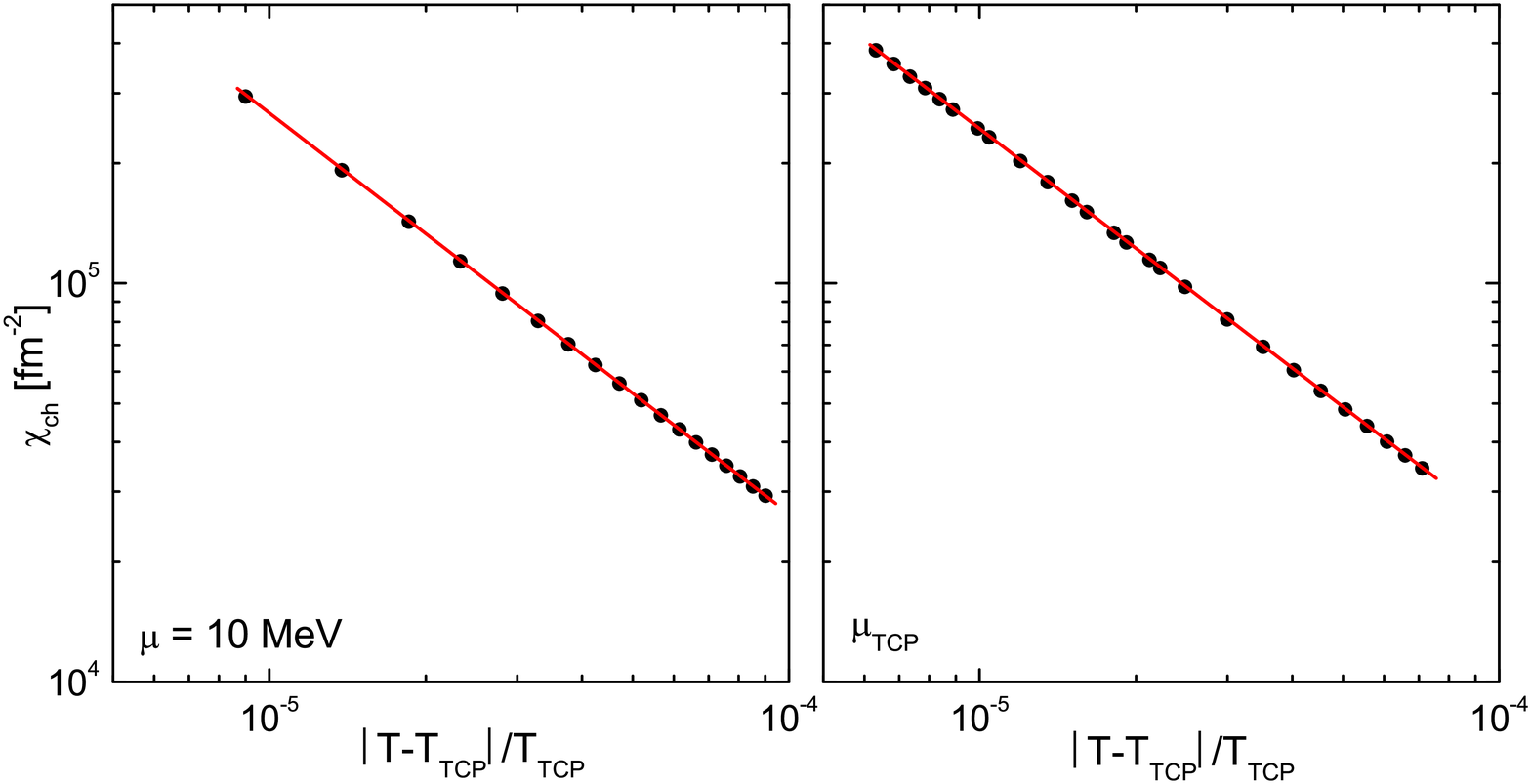}
\caption{Dependence of the chiral susceptibility $\chi_{ch}$ as a
function of $T$ for constant $\mu$ in the vicinity of an arbitrary
point (taken to correspond to $\mu=10$ MeV) in the 2nd order
transition line (left panel) and the TCP (right panel) in the
chiral limit for parametrization Set C.} \label{figTCP}
\end{figure}

As it is well known, in the region of second order phase transition
the behavior of several relevant thermodynamical quantities in the vecinity
of the phase transition is determined by the critical exponents.
In the case of the chiral and quark number susceptibilities,
$\chi_{ch}$ and $\chi_q$ respectively, and the specific heat $C_V$
they are usually defined by
\begin{eqnarray}
\chi_{ch}  =  |h - h_c|^{-\gamma_{ch}}
\qquad ; \qquad
\chi_{q}  =  |h - h_c|^{-\gamma_{q}}
\qquad ; \qquad
C_V  =  |h - h_c|^{-\alpha},
\label{critexp}
\end{eqnarray}
where $|h - h_c|$ is the distance to the critical point in the $(\mu, T)$ plane. Note that in the
chiral limit only trajectories approaching the transition from the chirally broken phase are relevant.
In the present case, given the Landau expansion obtained above, one expects to have the usual mean
field exponents.
For trajectories which are not asymptotically tangential to the critical line they are
\begin{equation}
\gamma_{ch}=1 \qquad ; \qquad \gamma_{q}=0\qquad ; \qquad \alpha =0,
\end{equation}
for all points except for the TCP where
\begin{equation}
\gamma_{ch} = 1 \qquad ; \qquad \gamma_{q}=1/2 \qquad ; \qquad \alpha =1/2.
\end{equation}
As a test of consistency we have determined them numerically by studying the asymptotic behavior
of the corresponding quantities for our three set of parameters. As an example of a typical result
of such studies we show in the left panel of Fig.\ref{figTCP} the behavior of $\chi_{ch}$ for Set C as we approach
an arbitrary point in second order transition line, i.e. a point {\it  different} from the TCP,
at constant $\mu$ (for definiteness we consider $\mu=10$ MeV). The right panel in Fig.\ref{figTCP} displays
the results of a equivalent study for $\mu = \mu_{TCP}$. The values of the critical exponents
extracted from this type of analysis for Set C are given in Table \ref{tab3}. Very similar results are
found for Set A and Set B. We see that in all cases the numerically obtained values are in very good agreement
with the mean field ones given above.

We finish this section by clarifying the role played by Polyakov
loop in enhancing the critical temperature at a given value of $\mu$,
at least
in the region where the transition is of second order. For
simplicity we consider the parametrization Set A where there is no
wave function renormalization. The condition $A_c=0$ implies $1/8G
= S_{21}$ (see App. A). However, following similar steps as those
described in App.B of Ref.\cite{Tum:2005}, it is possible to show
that for $T,\mu << \Lambda_0$ one has
\begin{equation}
S_{21} \simeq S_{21}^{app} = \frac{1}{8 \pi^2} \left( \frac{\Lambda_0^2}{4} - \left[ \frac{\pi^2}{3} -
\frac{2}{3} \left( \arccos \left[\frac{3 \Phi -1}{2}\right] \right)^2 \right] T^2 - \mu^2 \right).
\label{approx}
\end{equation}
In fact, we have checked numerically that in the relevant region
$T \leq 210$ MeV and  $\mu \leq 50$ MeV, Eq.(\ref{approx}) is
verified with an accuracy higher than 15\% for $\Phi \leq 0.3$.
Therefore, the condition $A_c =0$ leads to
\begin{equation}
T_c(\mu) \simeq
\frac{T^{(pq)}_c(\mu)}{\sqrt{1-\frac{2}{\pi^2}
\left(\arccos\left[\frac{3 \Phi_c -1}{2}\right] \right)^2}},
\label{phigen}
\end{equation}
where
\begin{equation}
T^{(pq)}_c(\mu) =
\frac{\sqrt{3}\Lambda_0}{2\pi} \sqrt{ 1 - \frac{4\pi^2}{3 G_S\Lambda_0^2} -
\frac{4\mu^2}{\Lambda_0^2} }.
\label{phi1}
\end{equation}
$T^{(pq)}_c(\mu)$ provides a good approximation to the critical temperature corresponding
to pure quark (pq) non-local model, i.e. the model with no coupling to the PL,
for the exponential regulator considered in parametrization Set A (see
Ref.\cite{Tum:2005} for details). Of course, in the presence of
PL-quark interactions the value of $\Phi_c$ in
Eq.(\ref{phigen}) has to be obtained by simultaneously solving the
corresponding gap equation, i.e. the second equation in
Eq.(\ref{othereqs}) in the present case. However, we clearly see
that for any value of $\Phi_c < 1$ we have $T_c(\mu) >
T^{pq}_c(\mu)$. For example, at $\mu=0$ one typically has $\Phi_c\simeq 0.2$
which implies $T_c(0)/ T^{pq}_c(0) \approx
1.66$. Since for the parametrization Set A we have $T^{pq}_c(0)=126$
MeV for the pure quark non-local model in the chiral limit, we see
that the coupling to the PL is expected to raise this value up to
$T_c \sim 209$ MeV which is in very good agreement with the
numerically found value listed in Table \ref{tab2}.
As it is clear from Eqs.(\ref{phigen},\ref{phi1}) a similar enhancement of
the critical temperature can be obtained at (low) finite $\mu$. On the other hand,
it should be noticed that in order to apply the present type
of analysis to relate the values of $\mu_c(T)$ predicted in models with and
without PL one must have a common range of temperatures for which the transition
is of second order. However, for the parametrizations considered here
this is not possible since they always lead to $T_{TCP} > T^{pq}_c(0)$.
For example, from Table II we see that Set A leads to $T_{TCP}=204.8$ MeV
to be compared with the value $T^{pq}_c(0)=126$ MeV quoted above.

\section{Phase diagram for finite quark mass}

\begin{figure}[hbt]
\includegraphics[width=1.0\textwidth]{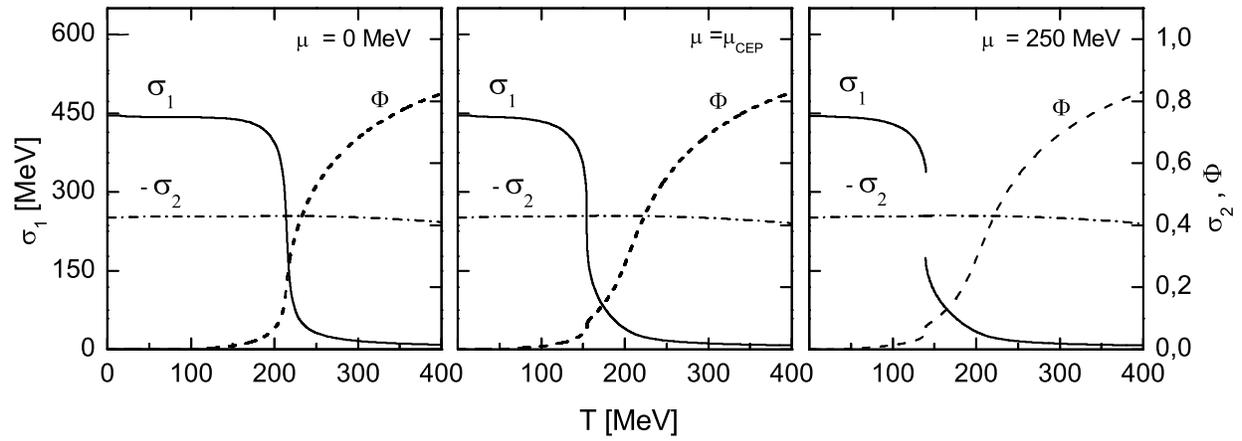}
\caption{Mean fields $\sigma_1$, $\sigma_2$ and $\Phi$ as functions of $T$ at three representative values
of chemical potentials for parametrization Set C.
Note that the scale to the left corresponds to $\sigma_1$
while that to the right to $\sigma_2$ and $\Phi$. Since $\sigma_2$ turns out
to be negative we plot $- \sigma_2$.}
\label{figMF}
\end{figure}

We start by analyzing the behavior of some mean field quantities
as functions of $T$ and $\mu$. Since the results obtained for our
three different parameterizations are qualitatively quite similar
we only present explicitly those corresponding to the
parametrization Set C. They are given in Fig.\ref{figMF} where we plot $\sigma_1$ , $\sigma_2$ and $\Phi$ as functions of $T$
for some representative values of the chemical potential. The left panel of Fig.\ref{figMF} shows that at
$\mu = 0$ there is a certain value of $T$ at which $\sigma_1$
drops rapidly signalling the existence of a chiral symmetry
restoration crossover transition. At basically the same
temperature the Polyakov loop $\Phi$ increases which can be
interpreted as the onset of the deconfinement transition. As $\mu$
increases there is a certain value of $\mu=\mu_{CEP}$ above which
the transition starts to be discontinuous. At this precise
chemical potential the transition is of second order. This
situation is illustrated in the central panel of Fig.\ref{figMF}. The corresponding values
$(T_{CEP}, \mu_{CEP})$ define the position of the so-called
``critical end point'' (CEP) which, as explained in App. B,
can be found
by solving a system of equations formed by the gap equations
Eqs.(\ref{fullgeq}) supplemented by two additional equations of the
type of Eqs.(\ref{second},\ref{third}).
As displayed
in the right panel of Fig.\ref{figMF}, for
$\mu > \mu_{CEP}$ the transition becomes discontinuous, i.e. of first
order. Finally, for chemical potentials above $\mu_c(T=0)\simeq
310$ MeV the system is in the chirally restored phase for all
values of the temperature. It is important to note that although
$\sigma_2$ appears to be rather constant in Fig.\ref{figMF}, at higher
values of $T$ it does go to zero as expected.

\begin{figure}[hbt]
\includegraphics[width=0.7\textwidth]{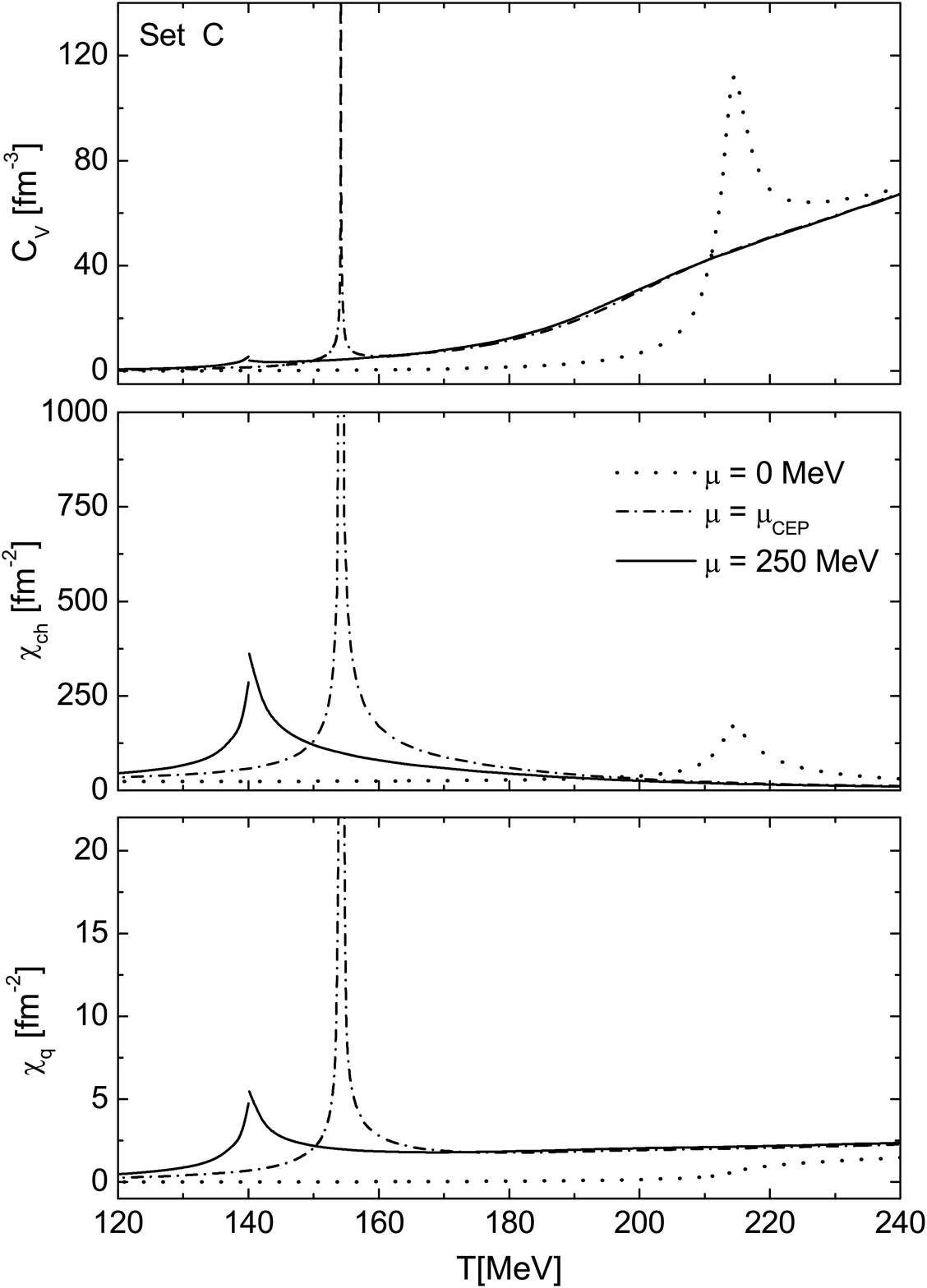}
\caption{Behavior of the specific heat $C_V$, the chiral
susceptibility $\chi_{ch}$ and the quark number susceptibility
$\chi_q$ as functions of $T$ at three representative values
of chemical potentials for parametrization Set C.} \label{figRespFun}
\end{figure}

The different nature of the chiral transition in each of the
three regions of Fig.\ref{figMF} is even more clearly observed in the behavior
of the corresponding response functions. In Fig.\ref{figRespFun} we display
the specific heat $C_V$ as well as the chiral and quark number susceptibilities,
$\chi_{ch}$ and {\bf $\chi_q$}, as a function of the temperature for parametrization Set C
and the three different values of $\mu$ used in Fig.\ref{figMF}.
The dotted line corresponds to $\mu = 0$. We observe that all the
response functions show a rather broad peak of finite height at
basically the same value of $T$. Such a value of $T$ corresponds
to the temperature at which the crossover transition
occurs. The dashed-dotted line corresponds to $\mu=\mu_{CEP}$ which
indicates that all the response functions display a sharp
and narrow divergent peak. Such a behavior signals the second
order nature of the chiral transition at the CEP. Finally, the
full line corresponds to $\mu > \mu_{CEP}$. In all cases we observe
a discontinuity in the response functions which indicates that
the associated transition is of first order.

\begin{figure}[hbt]
\includegraphics[width=0.9 \textwidth]{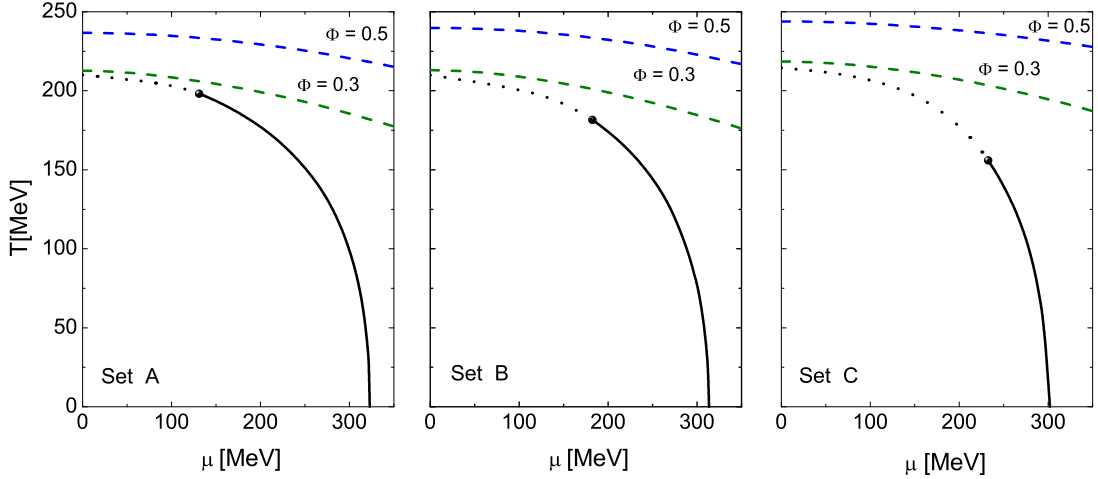}
\caption{Phase diagrams for the three parameterizations
considered. Set B and Set C include quark wave function renormalization
while Set A does not. Set A and Set B correspond to exponential form
factors while Set C to lattice motivated form factors. The dotted
line corresponds to the line of crossover chiral transition and
the full line to that of first order chiral transition. The dashed
lines correspond to the deconfinement transition (the lower and
upper lines being for $\Phi = 0.3$ and $\Phi = 0.5$,
respectively). } \label{figPhDiagFinM}
\end{figure}

\begin{figure}[hbt]
\includegraphics[width=0.9 \textwidth]{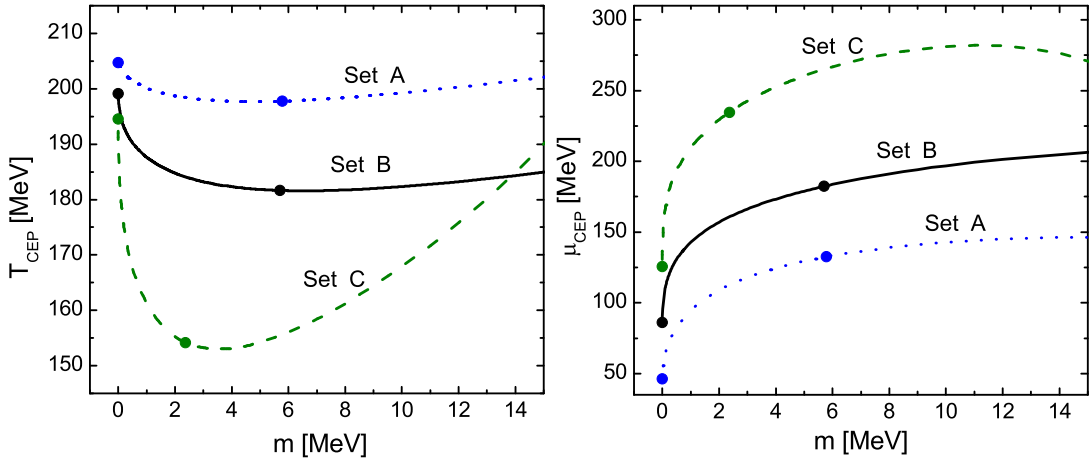}
\caption{Position of the CEP in the $(T,\mu)$ plane as a function
of the current quark mass $m$ for the three model
parameterizations used in this work. The left panel displays the
dependence of $T_{CEP}$ while the right panel that of $\mu_{CEP}$.
Values corresponding to the chiral limit $m=0$ and to the
``physical'' current quark masses given in Table I for each
parameter set are indicated by fat dot.  } \label{figCEPM}
\end{figure}

\begin{figure}[hbt]
\includegraphics[width=1.0 \textwidth]{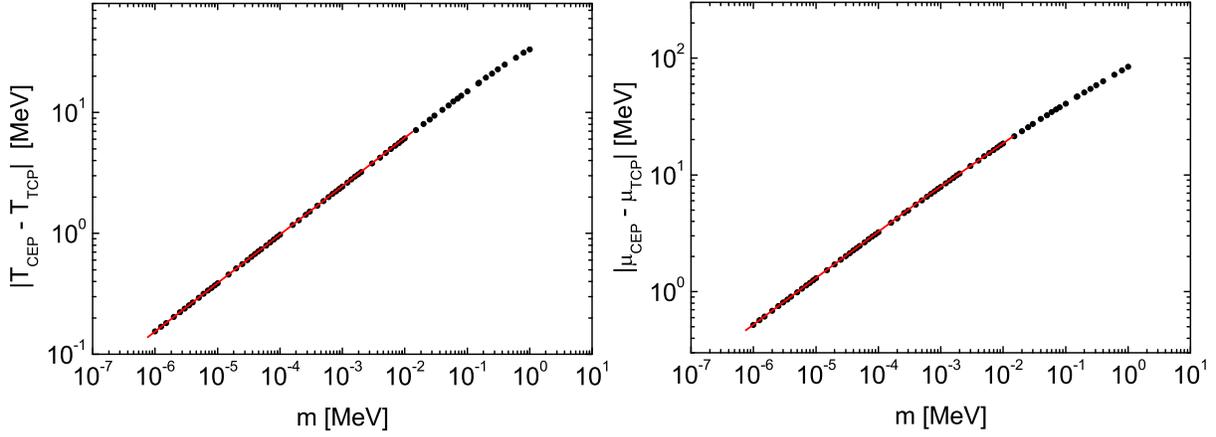}
\caption{Position of the CEP in the $(T,\mu)$ plane as a function
of the current quark mass $m$ close to the chiral limit for parametrization
Set C.}
\label{figexpCEP}
\end{figure}

The phase diagrams corresponding to our three different
parameterizations are given in Fig.\ref{figPhDiagFinM}. Here the
dotted line represents the line of crossover chiral transition
while the full line that of first order. The dashed lines are
associated to the deconfinement transition (the lower and upper
lines correspond to $\Phi = 0.3$ and $\Phi = 0.5$,
respectively). The position of the most relevant points in the
phase diagrams are tabulated in Table \ref{tab4}. As in the chiral
case, we observe that the main difference appears in the position
of the point at which the first order transition line ends.
Comparing the results of Set A and Set B we see that the main effect of
the wave function renormalization term is to shift the location of
the CEP towards lower values of $T$ and higher values of $\mu$.
Concerning the lattice motivated parametrization Set C we observe
that it leads to even lower values of $T_{CEP}$ and higher values
$\mu_{CEP}$. Comparing with the results obtained in the chiral
limit we note that the variation of $T_{CEP}$ between different
sets is larger in this case. Moreover, considering each
parametrization separately we observe that while the effect of
introducing a finite quark mass on the values of the $T_c(0)$ and
$\mu_c(0)$ is quite small (aprox. $+5$ MeV and $+30$ MeV,
respectively), the position of the CEP is much more sensitive to
the value of the current quark mass $m$, especially in the case of
parametrization Set C. This is clearly seen in Fig.\ref{figCEPM}
where we plot $T_{CEP}$ and $\mu_{CEP}$ as a function of $m$. In
all the cases we note a sharp decrease (increase) of $T_{CEP}$
($\mu_{CEP}$) for low values of $m$. In the case of the
exponential parametrization Set A and Set B, at $m \sim 4$ MeV this
variation tends to disappear and the position of the CEP remains
rather stable up to the maximum value of $m$ we have considered.
On the other hand, for the lattice motivated
parametrization the situation is somewhat different. In fact, the
variation is rather large for basically all the values of $m$
considered. In particular we see that, after the initial sharp
decrease, at about $4$ MeV the value of  $T_{CEP}$ starts to
increase in a rather pronounced way. It is interesting to note that
the behavior of the position of the CEP as a function of $m$ close
to $m=0$ (i.e. close to the TCP) can be analytically investigated
\cite{Hatta:2002sj}. General arguments imply that
\begin{eqnarray}
\Delta T_{CEP} &=& T_{CEP}(m) - T_{TCP} = - c \ m^{2/5} + {\cal O}(m^{4/5}) \nonumber
\\
 \Delta \mu_{CEP}&=& \mu_{CEP}(m) - \mu_{TCP} = + d \ m^{2/5} + {\cal O}(m^{4/5}),
\label{mexp}
\end{eqnarray}
where $c$ and $d$ are definite positive constants. To
verify that our results do satisfy these relations we have
numerically study in detail the variation of $T_{CEP}$ and $\mu_{CEP}$
as a function of $m$ close to the chiral limit. Results for
the parametrization Set C are shown in Fig.\ref{figexpCEP} where
the power-law behavior of both $\Delta T_{CEP}$ and $\Delta\mu_{CEP}$
is clear seen as a straight line in the corresponding log-log plot.
Performing a linear fit we obtain that the slope of both straight
lines is $0.40\pm 0.01$ in very good agreement with the exponents
in Eq.(\ref{mexp}).

\begin{figure}[hbt]
\includegraphics[width=0.9 \textwidth]{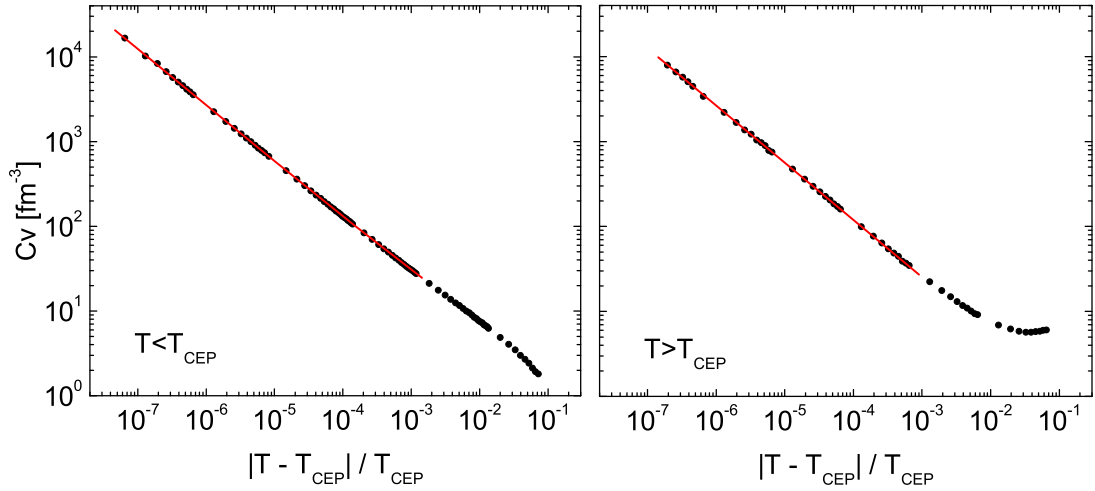}
\caption{Dependence of the specific heat $C_V$ as a function of
$T$ for constant $\mu$ in the vicinity of the CEP for
parametrization Set C. The left panel displays the dependence as
one approaches the CEP from below while the right panel the one
when the approach is done from above.} \label{figCEP}
\end{figure}

Finally, we consider the behavior of the response functions close
to the CEP. As already mentioned, the chiral phase transition at
this point is of second order. Thus, a critical behavior with
critical exponents defined as in Eq.(\ref{critexp}) is expected.
Within the approximations used in this work these exponents should
take the mean field values $\gamma_{ch}=\gamma_{q}=\alpha=2/3$
when one approaches the CEP using trajectories which are not
asymptotically parallel to the first order transition line. As
mentioned in Sec.III, these exponents can be numerically obtained
by analyzing the asymptotic behavior of the response functions
close to the critical point. In Fig.\ref{figCEP} we show a typical
result of this type of numerical study. There, we display a
log-log plot of the specific heat $C_V$ corresponding to the
parametrization Set C as a function of the relative temperature
departure $|T-T_{CEP}|/T_{CEP}$ for trajectories that approach the
CEP at constant $\mu = \mu_{CEP}$ both from below (i.e. $T <
T_{CEP}$) and from above (i.e. $T > T_{CEP}$). We observe that a
single straight line behavior is obtained up to relative departure
as large as $10^{-3}$ after which non-linear effects start to show
up. This result is particularly interesting since there have been
claims\cite{Hatta:2002sj,Costa:2007} that in some cases there might be a ``two
straight lines'' behavior, namely, that two different critical
exponents are needed in order to describe the critical behavior of
the $C_V$ close to the CEP. In fact, this feature has been
interpreted as an influence of the TCP critical properties on the
CEP ones. As displayed in Fig.\ref{figCEP} no sign of this type of
effect is found in our case. Results similar to those presented in
this figure have been obtained for the chiral and quark number
susceptibilities. The corresponding values of the critical
exponents for different kind of trajectories are listed in Table
\ref{tab8}. Note that in all cases the agreement with the mean
field values of the exponents is very good. The same type of
results has been found for the other two parameterizations, i.e.
the exponential parameterizations Set A and Set B. It should be
stressed that in order to obtain  numerically the critical
exponents with a good accuracy it is important to know the
position of the CEP with very good precision. We have checked that
even an error of 0.5 MeV in $T_{CEP}$ and/or $\mu_{CEP}$
translates into an important uncertainty in the critical
exponents. In this sense, the method for the determination of the
CEP discussed in App.~B is of great help.

\section{Summary and conclusions}

In this work we have studied the behavior of strongly interacting
matter at finite temperature and chemical potential using a
non-local chiral quark model which includes wave function
renormalization and coupling to the Polyakov loop. This type of
model can be understood as a non-local extension of the local
Polyakov-Nambu-Jona-Lasinio model, and represents a step towards a
more realistic modeling of the QCD interactions that could allow
a simultaneous description of the deconfinement and chiral phase
transition. The non-local interactions have been described by
considering not only the frequently used exponential form factors,
but also a parametrization based on fits to the quark mass and
renormalization functions obtained in lattice calculations. In
this framework, we have studied the corresponding phase diagrams
and associated quantities, both in the chiral limit and at finite
values of the current quark mass, paying particular attention to
the accurate determination of the critical points. In fact, in
both cases we have been able to obtain a set of coupled equations
for the position of the corresponding critical point, i.e. the
Tricritical Point (TCP) in the chiral limit and the Critical End
Point (CEP) for finite quark mass. Our numerical results indicate
that some of the features of the phase diagrams are not very much
dependent on the different parameterizations
we used. For example, for finite quark mass the
critical temperatures at $\mu=0$ are within the range $210-215$
MeV, while the critical chemical potentials at $T=0$ are in the
range of $298 - 322$ MeV. On the other hand, the position of the
critical point turns out to be very sensitive to both the
parametrization and the value of the current quark mass $m$.
Comparing the results corresponding to the exponential
parametrization with and without quark wave function
renormalization we find that the main effect of the presence of
this term is to shift the location of the CEP towards lower values
of $T$ and higher values $\mu$. Concerning the lattice motivated
parametrization we observe that it leads to even lower values of
$T_{CEP}$ and higher values $\mu_{CEP}$. As for the dependence on
$m$ we have verified that for small values of $m$ (i.e. close to
the TCP) the position of the CEP displays in all cases a power-law
behavior, as expected. For the exponential parameterizations at $m
\sim 4$ MeV this initial variation tends to disappear and the
position of the CEP remains rather stable up to the maximum value
of $m$ we have considered. On the other hand, for the lattice motivated parametrization the
situation is somewhat different. In fact, the dependence on $m$ is
rather strong for basically all the values considered. In
particular, after an initial sharp decrease, at $m \sim 4$ MeV the
value of $T_{CEP}$ starts to increase in a rather
pronounced way. Finally, we have analyzed
numerically the critical behavior around the TCP and the CEP
determining the critical exponents associated with the chiral and
quark number susceptibilities as well as the heat capacity. In all
cases, we find that the obtained exponents agree with
their predicted mean field values to a rather good degree of
accuracy. In particular, no influence of the TCP properties on the
CEP critical exponents has been found.

\section*{Acknowledgements}

We would like to acknowledge useful discussions with D. Gomez Dumm.
This  work was supported by CONICET (Argentina) grant \# PIP 00682
and by ANPCyT  (Argentina) grant \# PICT07 03-00818.

\section*{APPENDIX A: Derivation of the Landau expansion}

\newcounter{erasmo}
\renewcommand{\thesection}{\Alph{erasmo}}
\renewcommand{\theequation}{\Alph{erasmo}\arabic{equation}}
\setcounter{erasmo}{1} \setcounter{equation}{0} 

To derive the Landau expansion Eq.(\ref{expansion2}) we start by
assuming that the chemical potential $\mu$ is such that, in the
chiral limit, the chiral condensate vanishes at a critical
temperature $T_c(\mu)$. Since in that situation the mean field
value $\sigma_1$ also vanishes, for $T \sim T_c(\mu)$ it is
possible to perform a double expansion of $\Omega^{\rm
MFA}_{reg}$, Eq.(\ref{omegareg}), in powers of $\sigma_1$
and $m$. We obtain
\begin{eqnarray}
\Omega_{reg}^{\rm MFA}(\mu, T, \Phi, \sigma_2, \sigma_1)
& = &
\hat\Omega(\mu, T,\Phi, \sigma_2)~ + ~
4 \left[\frac{1}{8 G} - S_{21}(\mu, T,\Phi, \sigma_2)\right] \ \sigma_1^2
 + 2 ~S_{42}(\mu, T,\Phi, \sigma_2) \ \sigma_1^4
\nonumber \\
& &- 8~m \ \sigma_1
\left[ S_{11}(\mu, T,\Phi, \sigma_2) -
       S_{32}(\mu, T,\Phi, \sigma_2)\ \sigma_1^2 \right]~+~
{\cal O}(\sigma_1^6, m \ \sigma_1^5, m^2 \ \sigma_1^2)\ ,
\nonumber
\\
& &
\label{1stexp}
\end{eqnarray}
where
\begin{equation}
S_{jk}(\mu, T,\Phi, \sigma_2)\
=
\sum_{c} \int_{p,n}  g^{j}(\rho^{c} _{n,\vec{p}})
\left(\frac{Z(\rho^{c} _{n,\vec{p}})^2}{(\rho^{c}_{n,\vec{p}})^{2}}\right)^k \ ,
\end{equation}
and $\hat\Omega(\mu, T,\Phi, \sigma_2)$ is the MFA thermodynamical potential in the chiral limit
for vanishing $\sigma_1$. Namely,
\begin{eqnarray}
\hat\Omega(\mu, T,\Phi, \sigma_2) = \frac{8 T}{\pi^2}
\sum_{c} \int_{p,n} \mbox{ln} Z(\rho_{n, \vec{p}}^c) +
\frac{\kappa_p^2\ \sigma_2^2}{2\,G_S} + {\cal{U}}(\Phi
,T) + \Omega^{free}_{reg} + \Omega_0. \label{aa}
\end{eqnarray}
Using Eq.(\ref{cond})  the corresponding expression for the chiral condensate can be readily obtained. In the chiral limit we get
\begin{equation}
\langle\bar qq\rangle =
8  \ \sigma_1 \left[ S_{11}(\mu, T,\Phi, \sigma_2) - S_{32}(\mu, T,\Phi, \sigma_2) \sigma_1^2 \right] +
{\cal O}(\sigma_1^5).
\end{equation}
Inverting this equation and replacing in Eq.(\ref{1stexp}) we finally get
\begin{eqnarray}
\Omega_{reg}^{\rm MFA}(\mu, T, \Phi, \sigma_2, \langle \bar q q\rangle)
& = &
\hat\Omega(\mu, T,\Phi, \sigma_2)  +
A(\mu, T,\Phi, \sigma_2) \ \langle \bar q q\rangle^2  \nonumber \\
& + & C(\mu, T,\Phi, \sigma_2) \ \langle \bar q
q\rangle^4  + {\cal O}(\langle \bar q q\rangle^6) \
.\label{expansion}
\end{eqnarray}
Here, the coefficients $A$ and $C$ are given by
\begin{eqnarray}
A(\mu, T ,\Phi, \sigma_2)
& = & \frac{1}{4\, S_{11}^2(T,\mu,\Phi, \sigma_2)}\,
\left[\frac{1}{8G} - S_{21}(T,\mu,\Phi, \sigma_2)\right],
\nonumber \\
C(\mu, T ,\Phi, \sigma_2) & = & \frac{S_{42}(T,\mu,\Phi, \sigma_2)} {128\, S_{11}^4(T,\mu,\Phi, \sigma_2)}\
- \ \frac{S_{32}(T,\mu,\Phi, \sigma_2)}{32\,S_{11}^5(T,\mu,\Phi, \sigma_2)}\,
\left[\frac{1}{8G} - S_{21}(T,\mu,\Phi, \sigma_2)\right]\;.
\end{eqnarray}

The expansion Eq.(\ref{expansion}) looks very similar to Eq.(15) of Ref.\cite{Tum:2005}
where non-local
models in the absence of wave function renormalization and coupling to the Polyakov loop
were analyzed. In principle, following similar ideas, explicit equations for the second order
transition line as well as the position of the TCP point might be determined. However, the
fact that in the present case the coefficients $A$ and $C$ depend on the mean field values
$\sigma_2$ and $\Phi$ introduces further complications. In fact, slightly below
$T_c$ (for a given value of $\mu$) the non-vanishing value of the condensate induces departures
of $(\Phi, \sigma_2)$ from the critical values $(\Phi_c, \sigma_{2,c})$
obtained as solutions of the set of equations
\begin{eqnarray}
\frac{\partial\hat\Omega(\mu, T_c(\mu),\Phi, \sigma_2)}{\partial\sigma_{2}}\Big|_{\Phi_c, \sigma_{2,c}}=
\frac{\partial\hat\Omega(\mu, T_c(\mu),\Phi, \sigma_2)}{\partial\Phi}\Big|_{\Phi_c, \sigma_{2,c}}=0.
\label{othereqs}
\end{eqnarray}
Those departures can be obtained using the gap equations resulting
from Eq.(\ref{expansion}). To quadratic order in the condensate,
we obtain
\begin{eqnarray}
\sigma_2 &=&  \sigma_{2,c} - \frac{\left(
\partial^2_{\Phi \Phi}\hat\Omega_c\right)\left(
\partial_{\sigma_2} A_c\right) -
        \left( \partial^2_{\Phi \sigma_2}\hat\Omega_c\right) \left( \partial_{\Phi} A_c\right)}
        {\left( \partial^2_{\Phi \Phi}\hat\Omega_c\right) \left(\partial^2_{\sigma_2 \sigma_2}\hat\Omega_c\right)  -
         \left( \partial^2_{\Phi \sigma_2} \hat\Omega_c \right)^2 } \
\langle \bar q q\rangle^2,  \nonumber \\  \nonumber \\
\Phi &=&  \Phi_c - \frac{ \left(\partial^2_{\sigma_2 \sigma_2}\hat\Omega_c\right) \left(\partial_{\Phi} A_c\right) -
       \left(\partial^2_{\Phi \sigma_2}\hat\Omega_c\right) \left(\partial_{\sigma_2} A_c\right)}
       {\left( \partial^2_{\Phi \Phi}\hat\Omega_c\right) \left(\partial^2_{\sigma_2 \sigma_2}\hat\Omega_c\right)  -
         \left( \partial^2_{\Phi \sigma_2} \hat\Omega_c \right)^2 } \
       \langle \bar q q\rangle^2,
\label{expan}
\end{eqnarray}
where a compact notation has been used to denote the derivatives of $\hat\Omega$ and $A$
evaluated at $T_c$. In this notation we have, for example,
\begin{eqnarray}
\frac{\partial\hat\Omega(\mu, T_c(\mu),\Phi, \sigma_2)}{\partial\bar\Phi}\Big|_{\Phi_c, \sigma_{2,c}}
 =
\partial_{\Phi}\hat\Omega_c
\qquad ; \qquad
\frac{\partial A(\mu, T_c(\mu),\Phi, \sigma_2)}{\partial \sigma_2}\Big|_{\Phi_c, \sigma_{2,c}}
 =
\partial_{\sigma_2}A_c.
\end{eqnarray}
Using Eqs.(\ref{expan}) we can now obtain the leading corrections to $A$, $C$ and $\hat \Omega$ induced by the non-vanishing
value of the condensate. Replacing the corresponding results in Eq.(\ref{expansion}), and grouping in powers of
$\langle \bar q q\rangle$ we finally obtain Eq.(\ref{expansion2}). Namely,
\begin{eqnarray}
\Omega_{reg}^{\rm MFA} & = & \hat\Omega(\mu, T,\Phi_c, \sigma_{2,c})
+ A_c \ \langle \bar q q\rangle^2
+ C_c \ \langle \bar q q\rangle^4 + {\cal O}\left(\langle \bar q q\rangle^6\right),
\end{eqnarray}
where
\begin{eqnarray}
A_c & = & A(\mu, T_c(\mu),\Phi_c, \sigma_{2,c}),
\nonumber \\
C_c & = & C(\mu, T_c(\mu),\Phi_c, \sigma_{2,c}) - \nonumber \\
    & & \frac{
\left(\partial^2_{\sigma_2 \sigma_2}\hat\Omega_c\right)
\left(\partial_{\Phi}     A_c\right)^2 +
       \left(\partial^2_{\Phi     \Phi}    \hat\Omega_c\right) \left(\partial_{\sigma_2} A_c\right)^2 -
       2 \left(\partial^2_{\Phi \sigma_2}\hat\Omega_c\right) \left(\partial_{\sigma_2} A_c\right)
               \left(\partial_{\sigma_2} A_c\right)}
       {\left[ 2 \left( \partial^2_{\Phi \Phi}\hat\Omega_c\right) \left(\partial^2_{\sigma_2 \sigma_2}\hat\Omega_c\right)  -
         \left( \partial^2_{\Phi \sigma_2} \hat\Omega_c \right)^2
         \right]}.
\label{ab}
\end{eqnarray}

\section*{APPENDIX B: Formalism to determine the position of the CEP}

\newcounter{erasmo2}
\renewcommand{\thesection}{\Alph{erasmo2}}
\renewcommand{\theequation}{\Alph{erasmo2}\arabic{equation}}
\setcounter{erasmo2}{2} \setcounter{equation}{0} 

As discussed in Ref.\cite{Hatta:2002sj}, in the case in which the grand potential $\Omega$
depends on one single variational parameter, a set of equations that allows to determine
the position of the CEP can be obtained. This set is formed by the corresponding gap equation
supplemented with the two equations that result from demanding that the second and third
of grand potential with respect to the parameter also vanish. Physically, this corresponds to
determining the values of $(T, \mu)$ for which the grand potential around its
minimum is as flat as possible.
The purpose of this appendix is to give some details of the formalism needed
to generalize this idea to the case in which the grand potential depends on
more than one variational parameter. In
order to keep the derivation as general as possible we will assume here
that the grand potential depends on $N$ variational parameters
$\xi_1, \xi_2,..,\xi_N$.
For convenience, in what follows, we will distinguish $\xi_1$ from the rest
(note that there is no loss of generality in this choice since the ordering of the parameters
is completely arbitrary), and introduce the latin index $j=2,...,N$. Then, in a compact
notation, the set of gap equations reads
\begin{eqnarray}
\frac{\partial \Omega}{\partial \xi_1} & = & 0,
\label{gap1} \\
\frac{\partial \Omega}{\partial \xi_j} & = & 0.
\label{gapk}
\end{eqnarray}
The basic idea is now to consider the variational parameters $\xi_j$ as functions
of $\xi_1$ with the corresponding functional dependence determined by the solutions
of the $N-1$ gap equations Eqs.(\ref{gapk}). The total set of $N+2$ equations
needed to determine the CEP is then obtained by supplementing these $N-1$ equations with
those resulting from demanding that the first, second and third {\it total}
derivatives of the grand potential with respect to $\xi_1$ vanish. The first of these
equations turns out to be, of course, the gap equation Eq.(\ref{gap1}). The other two
result
\begin{eqnarray}
\frac{\partial^2 \Omega}{\partial \xi_1^2} +
2\ \frac{\partial^2 \Omega}{\partial \xi_1 \partial \xi_j}\ \xi'_j +
\frac{\partial^2 \Omega}{\partial \xi_j \partial \xi_k}\ \xi'_j \ \xi'_k
& = & 0,
\label{second}
\end{eqnarray}
\begin{eqnarray}
\frac{\partial^3 \Omega}{\partial \xi_1^3} +
3\ \frac{\partial^3 \Omega}{\partial \xi_1^2 \partial \xi_j}\ \xi'_j +
3\ \frac{\partial^3 \Omega}{\partial \xi_1 \partial \xi_j \partial \xi_k}\ \xi'_j \ \xi'_k & + & \nonumber \\
3\ \frac{\partial^3 \Omega}{\partial \xi_j \partial \xi_k \partial \xi_l}\ \xi'_j \  \xi'_k \ \xi'_l +
3\ \frac{\partial^2 \Omega}{\partial \xi_j \partial \xi_k}\ \xi'_j \ \xi''_k +
3\ \frac{\partial^2 \Omega}{\partial \xi_1 \partial \xi_j}\ \xi''_j
& = & 0.
\label{third}
\end{eqnarray}
Here, and in what follows, the sum over repeated latin indexes
$k,j,l=2,..,N$ is understood. Note that in obtaining these
equation some terms have been dropped assuming that the gap
equations Eqs.(\ref{gap1},\ref{gapk}) are simultaneously
satisfied. In Eqs.(\ref{second},\ref{third}), $\xi'_j$ ,
$\xi''_j$, etc., denote the derivatives of the corresponding
parameters with respect to $\xi_1$. They can be conveniently
expressed in terms of partial derivatives of the grand potential
by solving the two set of linear equations resulting from taking
the first and second total derivatives of both sides of the gap
equations Eq.(\ref{gapk}). We obtain
\begin{equation}
\xi'_j = - ( C^{-1} )_{jk} \ \frac{\partial^2 \Omega}{\partial \xi_1 \partial \xi_k},
\label{dos}
\end{equation}
\begin{equation}
\xi''_j = - ( C^{-1} )_{jk}
\left[ \frac{\partial^3 \Omega}{\partial \xi_1^2 \partial \xi_k}
+ 2 \frac{\partial^3 \Omega}{\partial \xi_1 \partial \xi_j \partial \xi_k}\ \xi'_k
+ \frac{\partial^3 \Omega}{\partial \xi_j \partial \xi_k \partial \xi_l}\ \xi'_k \ \xi'_l
\right],
\label{tres}
\end{equation}
where
\begin{equation}
C_{jk} = \frac{\partial^2 \Omega}{\partial \xi_j \partial \xi_k}.
\label{cuatro}
\end{equation}

Therefore, once the explicit form of the grand potential $\Omega$ in terms of the variational
parameters is known, all the derivatives appearing in the set of equations Eqs.(\ref{gap1}$-$\ref{third})
can be analytically determined. Then, the numerical solution of this set
of equations allows to determine the values of $(T, \mu)$ corresponding to the CEP as well as the
corresponding values of the variational parameters $\xi_1, \xi_2,..,\xi_N$.

Turning to the model discussed in the present work, we note that for the parameterizations Set B and Set C
we have three variational parameters. We chose to identify them as $\xi_1 = \sigma_1$,
$\xi_2 = \sigma_2$ and $\xi_3 = \Phi$. Then, the set of five equations needed to determine
the CEP is formed by Eqs.(\ref{fullgeq}) supplemented by the two equations that result from performing
the corresponding identifications in Eqs.(\ref{second}$-$\ref{cuatro}).
In the case of parametrization Set A the situation is somewhat simpler since there are only two variational
parameters ($\xi_1 = \sigma_1$ and $\xi_2 = \Phi$) and, hence, four equations are required to
determine the CEP.

\newpage

\begin{table}[h]
\caption{Set parameters and chiral condensates for $T=\mu=0$.}
\label{tab1}
\begin{center}%
\begin{tabular}
[c]{ccccc}\hline
&  & \hspace{.5cm} Set A \hspace{.5cm} & \hspace{.5cm} Set  B \hspace{.5cm} & \hspace{.5cm} Set C \hspace{.5cm}
\\\hline
$m_{c}$ & MeV & 5.78 & 5.70 & 2.37\\
$G_{s} \Lambda_{0}^{2}$ &  & 20.650  & 32.030 & 20.818\\
$\Lambda_{0}$ & MeV & 752.20 & 814.42 & 850.00\\
$\kappa_{P}$ & GeV & $-$ & 4.180 & 6.034\\
$\Lambda_{1}$ & MeV & $-$ & 1034.5 & 1400.0\\\hline
$\sigma_{1}$ & MeV & 424 & 529 & 442\\
$\sigma_{2}$ &  & $-$ & -0.43 & -0.43\\
$- <q\bar q>^{1/3}$ & MeV & 240 & 240 & 326\\\hline
\end{tabular}
\end{center}
\end{table}

\begin{table}[h]
\caption{Position of some characteristic points of the phase diagrams
in the chiral limit.  All values are in MeV.}
\label{tab2}
\begin{center}%
\begin{tabular}
[c]{ccccc}
\hline \hline
             & \hspace*{.5cm}   Set A \hspace*{.5cm}  &  \hspace*{.5cm}  Set B \hspace*{.5cm}  & \hspace*{.5cm}  Set C \hspace*{.5cm} \\
\hline
$T_c(0)$      & 206.6  &  205.9 & 209.7  \\
$\mu_{TCP}$   &  46.2  &   86.1 & 125.7  \\
$T_{TCP}$     & 204.8  &  199.2 & 194.6  \\
$\mu_c(0)$    & 297.6  &  285.5 & 268.2  \\
\hline \hline
\end{tabular}
\end{center}
\end{table}

\begin{table}[h]
\caption{Critical exponents in the chiral limit for Set C.} \label{tab3}
\begin{center}%
\begin{tabular}[c]{ccccccc}
\hline \hline
                     &$\ $&
   & \hspace*{.5cm}  $\gamma_{ch}$ \hspace*{.5cm}  & \hspace*{.5cm}  $\gamma_q$ \hspace*{.5cm}  &   \hspace*{.5cm}  $\alpha$  \hspace*{.5cm}   \\ \hline
Point in             &    & $\mu \rightarrow  $    &    1.00(1)     &     0.00(1)    &    0.00(1)    \\
   2nd order         &    & $ T \uparrow^{\! \! }$        &    1.00(1)     &     0.00(1)    &    0.00(1)    \\ \cline{3-6}
   critical line     &    & MF exponent          &      1       &      0       &     0       \\
\hline
                     &    & $\mu \rightarrow$      &    1.00(1)     &     0.51(1)    &    0.50(1)    \\
  TCP                &    & $ T \uparrow$          &    1.00(1)     &     0.51(1)    &    0.50(1)    \\ \cline{3-6}
                     &    & MF exponent           &      1       &      1/2    &     1/2       \\
 \hline \hline
\end{tabular}
\end{center}
\end{table}

\begin{table}[h]
\caption{Position of some characteristic points of the phase
diagrams for finite quark mass. All values are in MeV.}
\label{tab4}
\begin{center}%
\begin{tabular}
[c]{cccc} \hline \hline
              & \hspace*{.5cm} Set A \hspace*{.5cm}   & \hspace*{.5cm}    Set B \hspace*{.5cm}  & \hspace*{.5cm}  Set C \hspace*{.5cm} \\
\hline
$T_c(0)$       & 210.0 &  209.8 & 214.5  \\
$\mu_{CEP}$    & 132.5 &  182.3 & 234.8  \\
$T_{CEP}$      & 197.8 &  181.6 & 154.2  \\
$\mu_c(0)$     & 321.5 &  311.6 & 298.1  \\
\hline \hline
\end{tabular}
\end{center}
\end{table}

\begin{table}[h]
\caption{Critical exponents at the CEP for Set C.} \label{tab8}
\begin{center}%
\begin{tabular}[c]{ccccc}
\hline \hline
                  & \hspace*{.5cm}  $\gamma_{ch}$ \hspace*{.5cm}  & \hspace*{.5cm}  $\gamma_q$ \hspace*{.5cm}
                  & \hspace*{.5cm}           $\alpha$    \hspace*{.5cm}      \\
\hline
$\mu \rightarrow$ &  0.67(1)        &  0.67(1)           &        0.66(1)                  \\
 $\mu \leftarrow$ &  0.66(1)        &  0.66(1)           &        0.67(1)                  \\
$T \uparrow$      &  0.67(1)        &  0.67(1)           &        0.66(2)                  \\
 $T \downarrow$   &  0.66(1)        &  0.66(1)           &        0.67(1)                  \\
\hline
    MF exponent   &   2/3             &        2/3     &       2/3                   \\
\hline \hline
\end{tabular}
\end{center}
\end{table}

\vfill

\pagebreak

\end{document}